\documentclass[dvips,twoside,fleqn]{article}
\usepackage{times}
\usepackage{amstext}
\usepackage{amssymb}
\usepackage{espcrc2}
\usepackage{graphicx}
\title{Large rescaling of the Higgs condensate: theoretical 
motivations and lattice results}
\author{
P. Cea\address{INFN - Sezione di Bari - Via Amendola 173 - 70126 Bari - Italy},
M. Consoli\address{INFN - Sezione di Catania - Corso Italia 57 - 95129 Catania - Italy},
and L. Cosmai$^{\text{a}}$
}

\begin{document}

\begin{abstract}
In the Standard Model the Fermi constant is associated with the vacuum 
expectation value of 
the Higgs field , `the condensate', usually
believed to be a cutoff-independent quantity. 
General arguments related to the `triviality' of $\Phi^4$ theory 
in 4 space-time dimensions suggest,
however, a dramatic renormalization effect in the continuum limit that 
is clearly visible on the 
relatively large lattices available today. The result can be
crucial for the Higgs phenomenology and in any context 
where spontaneous symmetry breaking 
is induced through scalar fields.
\end{abstract}
\maketitle

%%%%%%%%%%%%%%%%%%%%%%%%%%%%%%%%%%%%%%%%%%%%%%%%%%%%%%%%%%%%%%%%%%%%%%%%%%%%%%%%%%%%%%%
\section{Introduction}
%%%%%%%%%%%%%%%%%%%%%%%%%%%%%%%%%%%%%%%%%%%%%%%%%%%%%%%%%%%%%%%%%%%%%%%%%%%%%%%%%%%%%%%

To understand
the scale dependence of the `Higgs condensate' 
$\langle\Phi\rangle$
let us define the 
$\lambda\Phi^4$ theory in the presence of a lattice 
spacing $a \sim 1/\Lambda$. 
The basic problem is the relation between
the bare ``lattice'' condensate
$v_B(\Lambda)\equiv \langle \Phi_{\rm latt} \rangle$  
and its renormalized physical value 
$v_R \equiv {{v_B  }\over{ \sqrt {Z } }}$
in the continuum limit $\Lambda \to \infty$.

In the presence of spontaneous symmetry breaking, 
there are two basically different definitions:
$Z\equiv Z_{\rm prop}$
from the propagator of the bare shifted `Higgs' field 
$h_B(x)\equiv\Phi_{\rm latt}(x)-v_B$
\begin{equation}
\label{prop}
       G(p^2) \sim { { Z_{\rm prop} }\over{p^2 + M^2_h} } 
\end{equation}
and $Z\equiv Z_\varphi$
where $Z_\varphi$ is the rescaling 
of $v^2_B$ needed in the effective potential 
$V_{\rm eff} (\varphi_B)$
\begin{equation}
\label{Zvarphi}
{\chi}^{-1}=
\left.   { { d^2V_ {\rm eff} } \over { d {\varphi}^2_B }} \right|_{ {\varphi}_B=v_B}\equiv
{{M^2_h}\over{Z_\varphi}}
\end{equation}
to match the quadratic shape at its absolute minima with the Higgs mass.
The usual assumption 
$Z_\varphi\sim Z_{\rm prop}$
is equivalent to require
a smooth limit  $p \to 0$. 
This is not necessarily true in the presence of
Bose condensation phenomena \cite{cs}
where one can have a very large particle density at
zero-momentum that compensates for the vanishing strength 
$\lambda \sim 1/\ln \Lambda$ of the elementary two-body processes.
In this case, one can have trivially free fluctuations 
so that $Z_{\rm prop} \to 1$ and
$h_B(x)=h_R(x)=h(x)$, 
but a non-trivial effective potential with 
a divergent 
$Z_\varphi\sim 1/\lambda \sim \ln {{\Lambda}\over{M_h}}$ \cite{cs}.
Therefore, 
the bare ratio 
$R_{\rm bare}= {{M^2_h}\over{v^2_B}} \to 0$
consistently with the rigorous 
results of Euclidean field theory~\cite{book} 
 but 
$R_\varphi= {{M^2_h}\over{v^2_R}}$
remains finite and cannot
be used to constrain the magnitude of $\Lambda$.

%%%%%%%%%%%%%%%%%%%%%%%%%%%%%%%%%%%%%%%%%%%%%%%%%%%%%%%%%%%%%%%%%%%%%%%%%%%%%%%%%%%%%%%
\section{The lattice simulation}
%%%%%%%%%%%%%%%%%%%%%%%%%%%%%%%%%%%%%%%%%%%%%%%%%%%%%%%%%%%%%%%%%%%%%%%%%%%%%%%%%%%%%%%

The one-component $(\lambda\Phi^4)_4$ theory   
becomes in the Ising limit
\begin{equation}
\label{ising}
S = -\kappa
\sum_x\sum_{\mu} \left[ 
\phi(x+\hat e_{\mu})\phi(x) +
\phi(x-\hat e_{\mu})\phi(x) \right]    
\end{equation}
with 
$\Phi(x)=\sqrt{2\kappa}\phi(x)$
 and where $\phi(x)$ takes only the 
values$+1$ or $-1$.  

We performed Monte-Carlo simulations of this Ising action 
using the Swendsen-Wang cluster algorithm.  
Lattice observables include:
the bare magnetization
$v_B=\langle |\Phi| \rangle$
where   $\Phi \equiv \sum_x \Phi(x)/L^4$
is the average field for each lattice configuration),
the zero-momentum susceptibility
$\chi=L^4 \left[ \left\langle |\Phi|^2 \right\rangle - 
\left\langle |\Phi| \right\rangle^2 \right]$,
the shifted-field propagator
\begin{equation}
G(p)= \langle \sum_x \exp (ip x) (\Phi(x) - v_B) 
(\Phi(0)- v_B) \rangle \, ,
\end{equation}
where $p_{\mu}={{2\pi}\over{L}}n_{\mu}$
with $n_{\mu}$
being a vector with integer-valued components, not all zero.

When approaching the continuum limit,
one can compare the lattice data for $G(p)$
to the 2-parameter formula 
\begin{equation}
\label{gpform}
G_{\mathrm{fit}}(p)= {{Z_{\rm prop}}\over{\hat{p}^2 + m^2_{\rm latt}} }
\end{equation}
where 
$m_{\rm latt}$
is the dimensionless lattice mass and 
$\hat{p}_{\mu}=2 \sin {{p_{\mu} }\over{2}}$. 
Actually, if ``triviality'' is true, 
there  must be a region of momenta
where Eq.(\ref{gpform}) gives a good description of the lattice data
and can be used to define the mass.
However, since the determination of the mass is a crucial issue, 
it is worth to compare the results of the previous method  
with the determination  in terms of ``time-slice'' 
variables~\cite{montweisz}.
To this end let us consider a lattice with 3-dimension 
$L^3$ and temporal 
dimension $L_t$ and the two-point correlator $C_1(t,0; {\bf k})$.
In this way, 
parameterizing the correlator 
$C_1$ in terms of the energy $\omega_k$, 
%as 
%\begin{equation}
%\label{fitcor}
%C_1(t,0;{\bf k})= A \, [ \, \exp(-\omega_k t)+\exp(-\omega_k(L_t-t)) \, ] \,,
%\end{equation}
the mass can be determined through the lattice dispersion relation~\cite{new}
\begin{equation}
\label{disp}
m^2_{\rm TS} = ~2 (\cosh \omega_k  -1)~~ -~~2 \sum ^{3} _{\mu=1}~ 
(1-\cos k_\mu) \,.
\end{equation}
%

%%%%%%%%%%%%%%%%%%%%%%%%%%%%%%%%%%%%%%%%%%%%%%%%%%%%%%%%%%%%%%%%%%%%%%%%%%%%%%%%%%%%%%%
\section{Numerical results: symmetric phase}
%%%%%%%%%%%%%%%%%%%%%%%%%%%%%%%%%%%%%%%%%%%%%%%%%%%%%%%%%%%%%%%%%%%%%%%%%%%%%%%%%%%%%%%

As a check of our simulations we started our analysis at  
$\kappa=0.0740$
in the symmetric phase, 
where the high-statistics results by Montvay \& Weisz~\cite{montweisz}
are available.   

Fig.~1 displays the data for the scalar propagator suitably re-scaled
in order to show the very good quality of the fit Eq.~(\ref{gpform}). 
The 2-parameter fit gives 
$m_{\mathrm{latt}}=0.2141(28)$ and 
$Z_{\rm prop} = 0.9682(23)$.
The value at zero-momentum is defined as  
$Z_\varphi \equiv m^2_{\rm latt} \chi = 0.9702(91)$.
Notice the perfect agreement between 
$Z_\varphi$ and  $Z_{\rm prop}$.
We measure also the time-slice mass Eq.~(\ref{disp}) 
at several
values of the  3-momentum. Our results are in 
good agreement with the corresponding result of 
Montvay \& Weisz~\cite{montweisz}
and with the value 
$m_{\rm latt} = 0.2141(28)$
obtained from the fit to the propagator data.
\begin{figure}[t]
\begin{center}
\includegraphics[width=0.45\textwidth,clip]{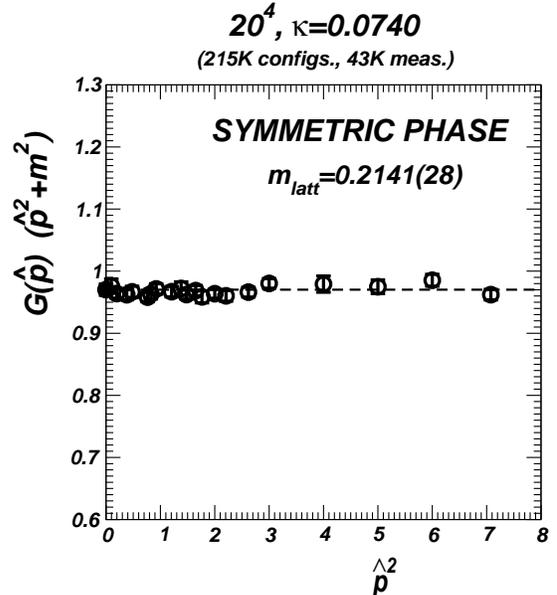}
\vspace{-1.24truecm}
\caption{The lattice data for the re-scaled propagator at $\kappa=0.0740$ in the
symmetric phase. The zero-momentum full point is defined as 
$Z_\varphi=m^2_{\mathrm{latt}}\chi$.
The dashed line indicates the value of $Z_{\mathrm{prop}}$.}
\end{center}
\end{figure}
%\begin{figure}[t]
%\begin{center}
%\includegraphics[width=0.45\textwidth]{figure_02.eps}
%\end{center}
%\end{figure}
In conclusion
our analysis of the 
symmetric phase is in  good agreement
with Ref.~\cite{{montweisz}} and supports, 
to high accuracy, 
the usual identifications 
$Z_\varphi \simeq Z_{\rm prop}$ and 
$m_{\rm latt} \simeq m_{\rm TS}$.
Note that our result for 
$Z_{\rm prop}$ is in excellent agreement with
the 1-loop renormalization group prediction~\cite{lw} 
$Z_{\rm pert} = 0.97(1)$.

%%%%%%%%%%%%%%%%%%%%%%%%%%%%%%%%%%%%%%%%%%%%%%%%%%%%%%%%%%%%%%%%%%%%%%%%%%%%%%%%%%%%%%%
\section{Numerical results: broken phase}
%%%%%%%%%%%%%%%%%%%%%%%%%%%%%%%%%%%%%%%%%%%%%%%%%%%%%%%%%%%%%%%%%%%%%%%%%%%%%%%%%%%%%%%

We now choose for $\kappa$  three successive values,
$\kappa = 0.076, 0.07512, 0.07504$, lying 
just above the critical $\kappa_c \simeq 0.0748$~\cite{montweisz}.
Thus, we are in the broken phase and 
approaching the continuum limit where the correlation length 
$\xi$ becomes 
much larger than the lattice spacing. 
To be confident that finite-size 
effects are sufficiently under control,
we used a lattice size, $L^4$, large enough so that 
$5 \lesssim L/\xi$~\cite{montweisz}.
We checked that or results for the magnetization and the susceptibility 
at $\kappa = 0.076$ are in excellent 
agreement with the corresponding results of Jansen {\it et al}~\cite{jansen}.  
Typical data for the re-scaled propagator are reported in Fig.~2.
Unlike Fig.~1, the fit to Eq.~(\ref{gpform}), though excellent at higher 
momenta, does not reproduce the lattice data down to zero-momentum. 
Therefore, in the broken phase, a meaningful determination of $Z_{\rm prop}$ 
and $m_{\rm latt}$ requires excluding the lowest-momentum points from the 
fit. 
The fitted $Z_{\rm prop}$ is slightly less than one.  This fact is 
attributable to residual interactions since we are not exactly at the 
continuum limit, so that the theory is not yet completely ``trivial.''  
This explanation is reasonable since we see a tendency for $Z_{\rm prop}$ 
to approach unity as we get closer to the continuum limit.  Moreover, 
we find good agreement between our result, $Z_{\mathrm{prop}}=0.9321(44)$, 
and the L\"uscher-Weisz perturbative prediction 
$Z_{\rm pert}=0.929(14)$~\cite{lw} at $\kappa = 0.0760$.  The comparison 
$Z_{\mathrm{prop}}=0.9566(13)$ with $Z_{\mathrm{pert}}=0.940(12)$ at 
$\kappa=0.07504$ is also fairly good.  
The quantity $Z_{\varphi}$ is obtained from the product 
$m_{\rm latt}^2 \chi$ and is shown in Fig.~3.  
According to conventional ideas $Z_{\varphi}$ should be {\it the same} as the 
wavefunction-renormalization constant, $Z_{\rm prop}$, but clearly it is 
significantly larger.  Note that there was no such discrepancy in Fig.~1 
for the symmetric phase. Moreover our data show that the discrepancy 
gets {\it worse} as we approach 
the critical $\kappa$.  Indeed Fig.~3 shows that $Z_{\varphi}$ grows rapidly 
as one approaches the continuum limit (where $m_{\rm latt} \to 0$).  
Thus, the effect cannot be explained by residual perturbative 
${\cal{O}}(\lambda_{\mathrm{R}}) $ effects that might cause $G(p)$ to 
deviate from the form in Eq.(\ref{gpform}); such effects die 
out in the continuum limit, according to ``triviality.''   
The results accord well with the ``two $Z$'' picture \cite{cs} in which, as 
we approach the continuum limit, we expect to see the zero-momentum point, 
$Z_\varphi \equiv m_{\rm latt}^2 \chi$, become higher and higher.
%
%Our  results for the magnetization
%and the susceptibility  are reported in Table 1. 
%\begin{table*}
%\tabcolsep 0.15cm
%\renewcommand{\arraystretch}{2}
%\begin{center}
%\begin{tabular}{ccclll}
%\hline \\[-0.95cm]
%\hline
%Ref.               &$L_{\text{size}}$  &$\#$ sweeps     &$\kappa$  
%&${{v_B}\over{\sqrt{2\kappa}}}$ &${{\chi}\over{2\kappa}}$      \\ \hline
%Our data             &$20^4$            &$5\times10^5$   &0.076 
%    &0.3015(1)         &37.71(22)          \\
%Jansen et al, 1989  &$20^4$  &$7.5\times10^6$ &0.076   
%  &0.30158(2)         &37.85(6)                \\
%Our data             &$32^4$            &$4\times10^5$   &0.07512    
% &0.1617(1)         &193.1$\pm$1.7                 \\
%Our data             &$32^4$            &$5\times10^5$   &0.07504   
%  &0.13822(12)         &293.38$\pm$2.86             \\
%\hline \\[-0.95cm]
%\hline
%\end{tabular}
%\caption{ The lattice data for the magnetization and the 
%susceptibility.}
% At $\kappa=0.076$ we report the corresponding results 
%obtained by Jansen et.al (Ref.\cite{jansen}).} 
%\end{center}
%\label{table:I}
%\end{table}
%
%\begin{figure}[t]
%\begin{center}
%\includegraphics[width=0.45\textwidth]{figure_03.eps}
%\end{center}
%\end{figure}
%\clearpage 
%\begin{figure}[t]
%\begin{center}
%\includegraphics[width=0.45\textwidth]{figure_04.eps}
%\end{center}
%\end{figure}
%\clearpage 
\begin{figure}[t]
\begin{center}
\includegraphics[width=0.45\textwidth,clip]{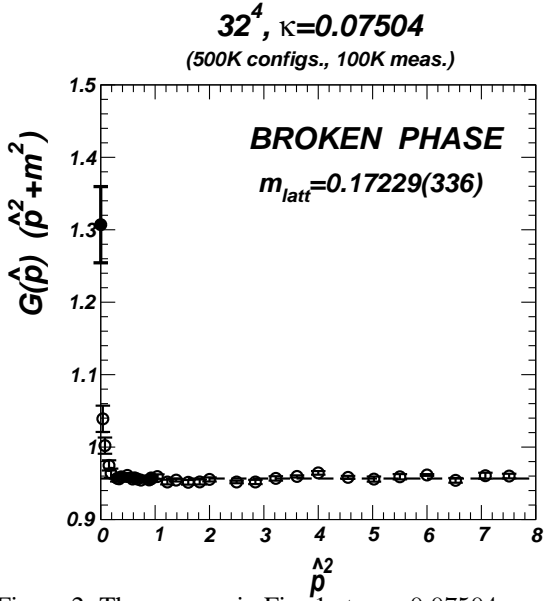}
\vspace{-1.24truecm}
\caption{The same as in Fig. 1 at $\kappa=0.07504$.}
\end{center}
\end{figure}
% 
% 
%We want to conclude by addressing a possible objection
%to our interpretation of the ``Higgs'' mass. 
%As explained, this is based on 
%the results of the
%fit to the finite-momentum propagator data.
%As an alternative strategy, one could decide to use only zero-momentum
%quantities relating  the Higgs mass to the value 
%$m_{\mathrm{TS}}(0)$  and
%defining $Z$ accordingly.
%
%\clearpage 
%\begin{figure}[t]
%\begin{center}
%\includegraphics[width=0.45\textwidth]{figure_06.eps}
%\end{center}
%\end{figure}
%\clearpage
\begin{figure}[t]
\begin{center}
\includegraphics[width=0.45\textwidth,clip]{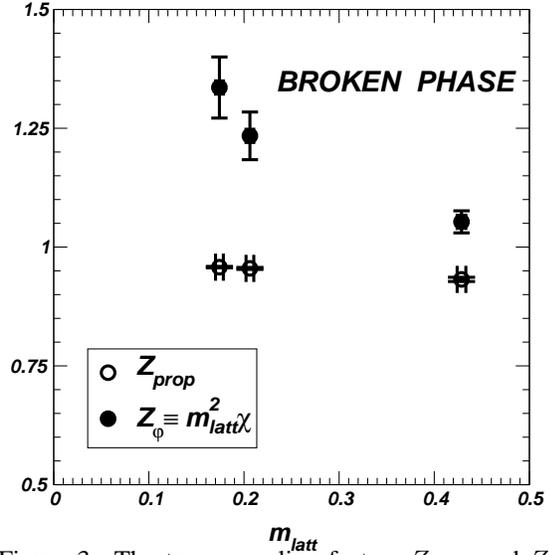}
\vspace{-1.24truecm}
\caption{The two re-scaling factors $Z_{\mathrm{prop}}$ and
$Z_\varphi$ as a function of $m_{\mathrm{latt}}$.  (The continuum limit 
corresponds to $m_{\rm latt} \to 0$.)}
\end{center}
\end{figure}
\section{Conclusions}
%%%%%%%%%%%%%%%%%%%%%%%%%%%%%%%%%%%%%%%%%%%%%%%%%%%%%%%%%%%%%%%%%%%%%%%%%%%%%%%%%%%%%%%

We have reported new numerical evidence that the 
re-scaling of the `Higgs condensate' 
$Z_{\varphi}\equiv m^2_{\rm latt} \chi$ 
is different from the conventional wavefunction renormalization 
$Z\equiv Z_{\rm prop}$.
Perturbatively, such a difference might 
be explicable if it became 
smaller and smaller when taking the continuum limit 
$\lambda_R \to 0$.  
However, our lattice data shows that the difference gets 
{\em larger} 
as one gets closer to the continuum limit, 
$m_{\rm latt} \to 0$.
Our lattice data is consistent with 
the unconventional picture \cite{cs} of 
``triviality'' and spontaneous symmetry breaking in which 
$Z_{\varphi}$ diverges 
logarithmically, while 
$Z_{\rm prop} \to 1$ in the continuum limit.  
In this picture the Higgs mass $M_h$ can remain finite 
in units of the 
Fermi-constant scale $v_R$, 
even though the ratio $M_h/v_B \to 0$.
The Higgs mass is then a genuine collective effect and 
$M_h^2$ is {\em not} 
proportional to the renormalized self-interaction strength.  If so, then 
the whole subject of Higgs mass limits is affected.

%%%%%%%%% Bibliography  %%%%%%%%%%%%%%%%%%%%%%%%%%%%%%%%%%%%%%%%%%%%%%%%%%%%%%%%%%%%


\vspace{-0.1cm}
\begin{thebibliography}{99}
\bibitem{cs}
M. Consoli and P. M. Stevenson, Zeit. Phys. C63 (1994) 427; Phys. Lett. 
B391 (1997) 144.
\bibitem{book} R. Fern\'andez, J. Fr\"ohlich, and A. D. Sokal, {\em
Random Walks, Critical  Phenomena, and Triviality in Quantum Field
Theory} (Springer-Verlag, Berlin, 1992).
%\bibitem{cs} M. Consoli and P. M. Stevenson, Zeit. Phys. {\bf C63}, 427 (1994).
%\bibitem{primer} M. Consoli and P. M. Stevenson, hep-ph/9407334.  
%\bibitem{response} M. Consoli and P. M. Stevenson, Phys. Lett. {\bf B391}, 
%144 (1997).
%\bibitem{old}  P. Cea, M. Consoli and L. Cosmai, Mod. Phys. Lett. 
%{\bf A13}, 2361 (1998).
\bibitem{montweisz} I. Montvay and P. Weisz, Nucl. Phys.  B290 (1987) 327.
\bibitem{new}  P. Cea, M. Consoli, L. Cosmai and P.M. Stevenson, 
Mod. Phys. Lett. A14 (1999) 1673.
\bibitem{lw} M. L\"{u}scher and  P. Weisz, Nucl. Phys. B 290 (1987) 25;
ibidem B295 (1988) 65.
\bibitem{jansen}
K. Jansen, I. Montvay, G. M\"unster, T. Trappenberg and U. Wolff, Nucl. Phys. 
B322 (1989) 698.  
\end{thebibliography}
\end{document}